\documentclass[11pt]{article}

\usepackage{amsmath,amssymb,amsthm,bm}
\usepackage[dvipdfmx]{graphicx}
\usepackage{psfrag,epsf}
\usepackage{enumerate}

\usepackage{natbib}
\newcommand{\blind}{0}

\usepackage{multicol}
\usepackage{flafter}
\usepackage{color,hyperref,xcolor}
\hypersetup{colorlinks=true,urlcolor=blue,citecolor=purple}
\usepackage{cleveref}
\numberwithin{equation}{section}

\newcommand{\abs}[1]{\left\lvert #1 \right\rvert}

\newcommand{\poi}{\mathrm{Poisson}}

\usepackage{adjustbox}
\usepackage{array}
\newcolumntype{R}[2]{%
    >{\adjustbox{angle=#1,lap=\width-(#2)}\bgroup}%
    l%
    <{\egroup}%
}
\newcommand*\rot{\multicolumn{1}{R{0}{2.5em}}}

\newtheoremstyle{general}
{3mm} 
{3mm} 
{} 
{} 
{\bfseries} 
{.} 
{.5em} 
{} 

\theoremstyle{general}
\newtheorem{assumption}{Assumption}
\newtheorem{theorem}{Theorem}
\newtheorem{corollary}{Corollary}
\newtheorem{definition}{Definition}
\newtheorem{remark}{Remark}

\begin{document}

\def\spacingset#1{\renewcommand{\baselinestretch}%
{#1}\small\normalsize} \spacingset{1}


\if0\blind
{
  \title{\bf A mathematical take on the competitive balance of a football league}
  \author{Soudeep Deb\thanks{Corresponding author. Email: soudeep@iimb.ac.in. ORCiD: 0000-0003-0567-7339}\hspace{.2cm}\\
    Indian Institute of Management, Bangalore \\ Bannerghatta Main Rd, Bilekahalli \\ Bengaluru, Karnataka 560076, India.}
  \maketitle
} \fi

\if1\blind
{
  \bigskip
  \bigskip
  \bigskip
  \begin{center}
    {\Large\bf A mathematical take on the competitive balance of a football league}
\end{center}
  \medskip
} \fi

\bigskip
\begin{abstract}
Competitive balance in a football league is extremely important from the perspective of economic growth of the industry. Many researchers have earlier proposed different measures of competitive balance, which are primarily adapted from the standard economic theory. However, these measures fail to capture the finer nuances of the game. In this work, we discuss a new framework which is more suitable for a football league. First, we present a mathematical proof of an ideal situation where a football league becomes perfectly balanced. Next, a goal based index for competitive balance is developed. We present relevant theoretical results and show how the proposed index can be used to formally test for the presence of imbalance. The methods are implemented on the data from top five European leagues, and it shows that the new approach can better explain the changes in the seasonal competitive balance of the leagues. Further, using appropriate panel data models, we show that the proposed index is more suitable to analyze the variability in total revenues of the football leagues.
\end{abstract}

\noindent%
\vfill

\newpage
\spacingset{1} 

\section{Introduction}
\label{sec:introduction}

Football (or association football, or soccer) is arguably the most popular sport in the world. Especially in Europe, it is not only popular, but also a big industry and is growing every year. In 2018/19 season, the total revenue in Europe's top five leagues was approximately 17 billion Euros, which is more than double of what it was a decade ago. \cite{buraimo2009market}, \cite{dima2015business} and \cite{rohde2017market} are some interesting reads on the football market in Europe. Now, with respect to the growth of the market, a common hypothesis is that the competitive balance in a sports league, which primarily relies on the strength and talent of the teams, is a crucial factor to affect public interest and thereby the financial health of the industry of that sports. \cite{rottenberg1956baseball} is a seminal work in this context. This study brought forth the Uncertainty of Outcome Hypothesis (UOH) in the sports economics literature. One can argue that if there is a high degree of uncertainty about the outcomes of a match, it attracts more viewers and thereby more revenue for the teams as well as the league. There have been a few interesting studies over the last couple of decades where the authors have analyzed the effect of competitive balance on attendance, broadcast rights or revenue sharing of the football industry in Europe. \cite{kesenne2000revenue}, \cite{szymanski2001income}, \cite{peeters2011broadcast} and \cite{plumley2018mind} are some notable examples.

Naturally, it is of primary importance to have a good measure of competitive balance. \cite{szymanski2003economic} distinguished among three different ways of measuring competitive balance: match-level, championship-level and season-level. A season-level measure is the most appropriate in capturing the competitiveness of a league as a whole. One of the earliest and standard techniques in this regard is to use the ratio of standard deviation of actual win percentages in the league to the same in an ideal league where each match can end in equally likely outcomes. While this measure and its variants have been used extensively in many studies (see \cite{quirk1997pay} and \cite{humphreys2002alternative} for example), \cite{michie2004competitive} argued in detail why it is not adequate in the context of football. One of the main limitations of this measure is that it was primarily developed for major sports in the United States of America (such as baseball, basketball etc) where draws are almost non-existent, but that is not the case for football. For instance, in the 2018-19 season, approximately 23.9\%, 18.7\%, 29\%, 29\% and 28.4\% matches were drawn respectively in Bundesliga, English Premier League (EPL), La Liga, Ligue 1 and Serie A. Another significant drawback of the above measure is that it fails to capture the presence of dominance in the league. In order to circumvent these issues, two other measures have been developed from the standard economic theory. First one is inspired from the concept of concentration ratio (see \cite{hall1967measures}) which quantifies the fraction of industry size held by the top $L$ firms. In case of a football league, one can consider the top $L$ clubs and compute the total proportion of their point share to define the concentration ratio. Note that the first four to six teams (depending on the league) of every league qualify for the European competitions. In view of that, we use $L=6$ and define the six club concentration index or C6 index as follows.

\begin{definition}[C6 index]
\label{def:c6index}
For a league with $n$ teams, let $S_i$ be the total points gathered by the $i$th team at the end of the season. The point share of the $i$th team is defined by $P_i = S_i/\sum_{i=1}^n S_i$. Let $P_{(j)}$ denote the $j$th largest value in the set $\{P_1,P_2,\hdots,P_n\}$. Then, the C6 index is defined as
\begin{equation}
    \label{eq:c6}
    \mathrm{C6} = \frac{n}{6}\sum_{j=1}^6 P_{(j)}.
\end{equation}
\end{definition}

The C6 index quantifies the imbalance between the top six clubs in a league and the rest, but it fails to capture the change in imbalance within these two groups. The second measure we shall discuss is more suitable in that aspect. It is derived from the Herfindahl-Hirschman index (HHI) and it captures the inequalities between all the clubs in a league. 

\begin{definition}[Herfindahl index of competitive balance]
\label{def:hicb}
Consider a league with $n$ teams, and let $P_i$ be the point share of the $i$th team as defined in \Cref{def:c6index}. Then, the Herfindahl index of competitive balance (HICB) is defined as
\begin{equation}
    \label{eq:hicb}
    \mathrm{HICB} = n\sum_{i=1}^n P_i^2.
\end{equation}
\end{definition}

In essence, HICB is a weighted average index for the league where the weights are proportional to each club's point share. Evidently, it reflects the extent of competitive balance among all the teams. As the competitive balance declines, there is a greater inequality among the teams and hence the value of either of the above two measures increases. The reader is referred to \cite{zimbalist2002competitive}, \cite{brandes2007made} and the references therein for more relevant discussions.

The above two measures have been widely used and can be considered to be benchmarks for measuring concentration of firms within an industry. While the adaptation to football leagues is natural and understandable, there is an interesting limitation that most authors have failed to address. In football leagues, it is not rare to have more than one team finishing up with same number of points, in which case the standing of those teams is usually determined by the overall goal differences, followed by other criteria such as goals scored, head-to-head record etc. Naturally, ultimate competitive balance should not only mean equally probable outcomes of a match, but it should also mean identically distributed goal differences (and goals scored) for the teams. Motivated by this, we propose a new measure in this paper which uses the actual scorelines of the games instead of just the outcomes. We also derive attractive theoretical properties of the proposed measure for a perfectly balanced league in the sense of the definition provided below. 

\begin{definition}[Perfectly balanced league]
\label{def:competitive-highest}
A league is `perfectly balanced' if any permutation of the teams is equally likely to be the final standing. 
\end{definition}

The paper is structured as follows. \Cref{sec:methods} lays out the theoretical background, along with relevant results and methods to analyze the data. A short simulation study is presented in \Cref{sec:simulation}. Next, we analyze a decade long data from the top five European leagues using the proposed method, and provide a comparative study of the three measures in terms of explaining the changes in overall revenues. These results are provided in \Cref{sec:results}. Following that, we summarize the advantages of the proposed approach in \Cref{sec:discussion}. Some important concluding remarks are also presented in that section.

\section{Methods}
\label{sec:methods}

In the discussions below, $n$ denotes the number of teams in the league. All of them play against each other in home and away system, implying that the total number of matches is $N=n(n-1)$. $Z_{ij1}$ (correspondingly, $Z_{ij0}$) is the outcome of the match between the $i$th and the $j$th team, played in the home ground of the $i$th team (correspondingly, the $j$th team). For instance, $Z_{ij1} = 1,0,-1$ indicate respectively a win, draw and loss at home for the $i$th team against the $j$th team. Further, let $S_{ij1}$ (and $S_{ij0}$) denote the points gathered by the $i$th team against the $j$th team at home (and at away); and let $X_{ij1}$ and $X_{ij0}$ be the corresponding number of goals scored. We shall use $S_i,GS_i,GC_i$ and $GD_i$ to denote the overall points, total goals scored, total goals conceded and the overall goal difference of the $i$th team. Thus,
\begin{equation}
\label{eq:overall-numbers}
    S_i = \underset{j\ne i}{\sum_{j=1}^n}\sum_{k=0}^1 S_{ijk}, \ GS_i = \underset{j\ne i}{\sum_{j=1}^n}\sum_{k=0}^1 X_{ijk}, \ GC_i = \underset{j\ne i}{\sum_{j=1}^n}\sum_{k=0}^1 X_{jik}, \ GD_i = GS_i - GC_i.
\end{equation}

Note that, for $i=1,2,\hdots,n$, the final standing of the league is determined first by $S_i$, then by $GD_i$ and then by $GS_i$. This is the standard norm for most domestic leagues and we shall consider this rule throughout the study. 

\begin{assumption}
\label{asmp:iid}
$X_{ijk}$, for all $i,j \in \{1,2,\hdots,n\}, k\in\{1,0\}$, are independent and $X_{ijk}$ follows a Poisson distribution with mean $\lambda_{ijk}$.
\end{assumption}

Our main theory revolves around the above assumption. Note that different values of $\lambda_{ijk}$ can be attributed to various factors that affect the number of goals scored by a team in a match. See \cite{clarke1995home} and \cite{everson2008composite} for related discussions. In what follows, first we present an ideal setup in which a league becomes perfectly balanced and then we develop an index of competitive balance based on how much a league deviates from that ideal setup. 

\begin{theorem}
\label{thm:lambda0-equal-prob}
If \Cref{asmp:iid} is true with $\lambda_{ijk}=\lambda>0$ for all $i,j \in \{1,2,\hdots,n\}, k\in\{1,0\}$, then there exists a $\lambda_0$ such that every match is equally likely to end in a win, draw or loss for any team.
\end{theorem}

Detailed proof of the above theorem is relegated to the Appendix for brevity. It leverages the concept of the modified Bessel function of the first kind. From the point of view of this paper, \Cref{thm:lambda0-equal-prob} provides an attractive ideal framework where a football league ensures ultimate competitive balance in the sense that every outcome of every match is equally likely. An immediate consequence is the following.

\begin{corollary}
If $X_{ijk}$, for all $i,j \in \{1,2,\hdots,n\}, k\in\{1,0\}$, are iid $\poi(\lambda_0)$, then a football league is perfectly balanced.
\end{corollary}

The proof is trivial in view of the fact that every match in the league is iid multinomial with the probability of every outcome being $1/3$ and that the total number of goals scored (and conceded) by every team are iid Poisson with mean $2\lambda_0(n-1)$. The latter ensures that in case two or more teams have equal number of points, they are equally likely to have a better goal difference (similarly, better goals scored, better head-to-head record as well) than others.

\begin{figure}[!htb]
    \centering
    \includegraphics[width = \textwidth,keepaspectratio]{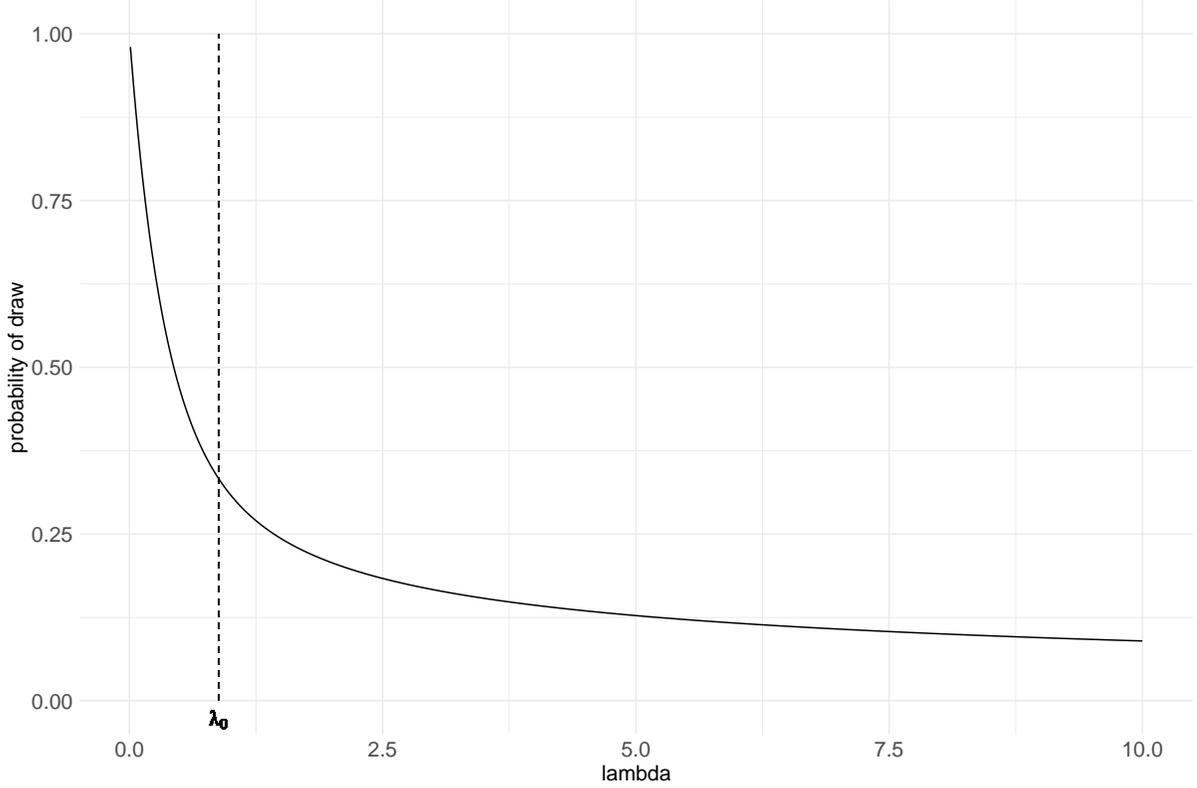}
    \caption{Probability of draw for different values of $\lambda$. Ideal situation of $\lambda_0\approx 0.88$ is displayed with a dotted line.}
    \label{fig:draw-lambda0}
\end{figure}

\begin{remark}
The value of $\lambda_0$ is computed to be approximately 0.88. \Cref{fig:draw-lambda0} further shows how the probability of draw changes with different values of $\lambda$. Note that, under the assumptions of \Cref{thm:lambda0-equal-prob}, if the probability of draw is $d_\lambda$, then any team can win or lose with equal probability $(1-d_\lambda)/2$. From the figure, it can be observed that as $\lambda\to 0$, a draw is expected whereas for large $\lambda$, the chance of having a definitive result increases.
\end{remark}

Interestingly, $\lambda_0$ is not the only value that makes a league perfectly balanced. It ensures that every outcome of a match is equally likely, a sufficient condition for a perfectly balanced league. However, this is not a necessary condition. In fact, the following is true. 

\begin{theorem}
\label{thm:lambda-competitive}
If every win, draw and loss are awarded 3, 1 and 0 points respectively, and if \Cref{asmp:iid} is true with $\lambda_{ijk}=\lambda>0$ for all $i,j \in \{1,2,\hdots,n\}, k\in\{1,0\}$, then a football league is perfectly balanced.
\end{theorem}

The above-mentioned point structure has been the practice in football leagues over the last few decades. A detailed proof is presented in the Appendix. We use properties of a Skellam distribution and some combinatorial arguments in the proof. We also point out that even if the point structure is different, similar idea can be used to get similar results.

Note that under the assumption that all $X_{ijk}$ are iid $\poi(\lambda)$, the maximum likelihood estimate (MLE) of $\lambda$ is given by
\begin{equation}
    \label{eq:lambda-mle}
    \hat\lambda = \frac{1}{2N} \underset{i\ne j}{\sum_{i=1}^n\sum_{j=1}^n} \sum_{k=0}^1 X_{ijk},
\end{equation}
which is essentially the average number of goals scored in the league. Based on $\hat\lambda$, one can also estimate the theoretical probability of draws under \Cref{asmp:iid}. Hereafter, it is denoted by $d_{\hat\lambda}$. In \Cref{sec:results}, the values of $\hat\lambda$ and $d_{\hat\lambda}$ for every league and every season in this study are presented. 

We next move on to developing a new measure of competitive balance. Recall that the mean equals the variance for a Poisson distribution. Under the general structure of \Cref{asmp:iid}, the $X_{ijk}$'s are not identically distributed and are dependent on individual team's skill, opponent, form or the ground. Therefore, the observed variance of the number of goals should be considerably more than the mean. In that light, if we assume that the $X_{ijk}$'s are independently distributed Poisson random variables, it is the deviations of $X_{ijk}$'s from $\hat\lambda$ which would speak about how much a league deviates from the hypothetical situation of being perfectly balanced. That motivates us to develop the following. 

\begin{definition}[Goal-based index]
\label{def:competitive-measure}
Let $X_{ij1}$ (and $X_{ij0}$) denote the number of goals scored by the $i$th team against the $j$th team at home (and at away). $\hat\lambda$ is the overall mean number of goals in the entire league. Then, the following goal-based index (GBI) is an appropriate measure of competitive balance of the league.
\begin{equation}
    \label{eq:competitive-measure}
    \mathrm{GBI} = \frac{1}{(2N-1)\hat\lambda} \underset{i\ne j}{\sum_{i=1}^n\sum_{j=1}^n} \sum_{k=0}^1 (X_{ijk}-\hat\lambda)^2.
\end{equation}
\end{definition}

\begin{theorem}
\label{thm:gbi-asymptotic}
Assume that $X_{ijk}$, for all $i,j \in \{1,2,\hdots,n\}, k\in\{1,0\}$, are independently distributed Poisson random variables. Then, for a league perfectly balanced in the sense of \Cref{thm:lambda-competitive}, $(2N-1)\mathrm{GBI}$ is asymptotically equivalent to a chi-squared random variable with degrees of freedom $2N-1$.
\end{theorem}

The proof is based on the theory of likelihood ratio tests and is presented in the Appendix. GBI, in comparison to C6 or HICB, is more attractive as it takes into account the number of goals scored by the teams, and not just the individual results of the matches. Moreover, \Cref{thm:gbi-asymptotic} provides a framework to formally test if a league shows significant deviation from being perfectly balanced or not. One can reject the null hypothesis of perfect balance if $(2N-1)GBI$ is greater than the critical value $\chi^2_{2N-1;\alpha}$, the upper $\alpha$-quantile of the corresponding chi-squared distribution. 

From a more applied point of view, it is worth checking whether GBI has better use over C6 or HICB in explaining the variation of revenue in a league. To that end, we collect the data on season-wide total commercial revenue of the five leagues (source: \cite{deloitterevenue}). Let $R_{lt}$ be the revenue (in billion Euros) of the $l$th league in the $t$th season. Let $\mathrm{CB}_{lt}$ be a measure of competitive balance for the same (we shall apply the following method for all of the above three measures). Our objective is to understand if $R_{lt}$ is significantly dependent on $\mathrm{CB}_{lt}$. The framework here is suitable for a panel data analysis, and hence, we use the following model.
\begin{equation}
    \label{eq:lm-revenue}
    R_{lt} = \alpha_l + \beta t + \gamma \mathrm{CB}_{lt} + \epsilon_{lt}, \; 1\le l\le 5, 1\le t\le T,
\end{equation}
where $\epsilon_{lt}$ are iid errors, $\alpha_l$ is the effect of the $l$th league, $\beta$ captures the effect of a linear trend and $\gamma$ is the main parameter of interest as it describes how the measure of competitive balance affects the revenue. For $\alpha_l$, we consider both fixed-effect and random-effect structures and compare their results. In case of random-effect models, $\alpha_l$ is taken as $\alpha_0 + \alpha_l'$ where $\alpha_0$ is the overall mean, and $\alpha_l'$ is the zero-mean random effect for the $l$th league. After fitting both types of models, we perform Hausman test (\cite{hausman1978specification}) to find out which one works better. 

All of the computations in this paper are done in RStudio version 1.2.5033, coupled with R version 3.6.2. The panel data analysis is carried out using the `plm' package by \cite{plmpackage}.

\section{Simulation study}
\label{sec:simulation}

We illustrate the theory from \Cref{thm:lambda0-equal-prob} and \Cref{thm:lambda-competitive} by simulating toy examples. For different values of $\lambda$, large number of leagues are generated under \Cref{asmp:iid}. For each league, we calculate the proportion of draws and plot them against the values of $\lambda$. Next, we look at the final standing of the league table and calculate the proportion of times each possible permutation is happening. Under the assumptions we have, \Cref{thm:lambda-competitive} suggests that the league is perfectly balanced and hence every permutation of the teams is equally likely to happen. That indicates a multinomial distribution with equal cell probabilities. On that note, using the observed frequencies of every permutation from the simulated data, we conduct multinomial goodness of fit test and find out the corresponding $p$-values. Note that the number of possible permutations for a league with $n$ teams is $n!$ which is quite large even for small $n$. So, in order to conduct an appropriate goodness of fit test, one needs to replicate the experiments millions of times. For computational ease and since the objective is to illustrate the results which are already proved, we limit ourselves to $n=5$ in this case and take 10000 replications in all experiments.

\begin{figure}[!hbt]
    \centering
    \includegraphics[width = \textwidth,keepaspectratio]{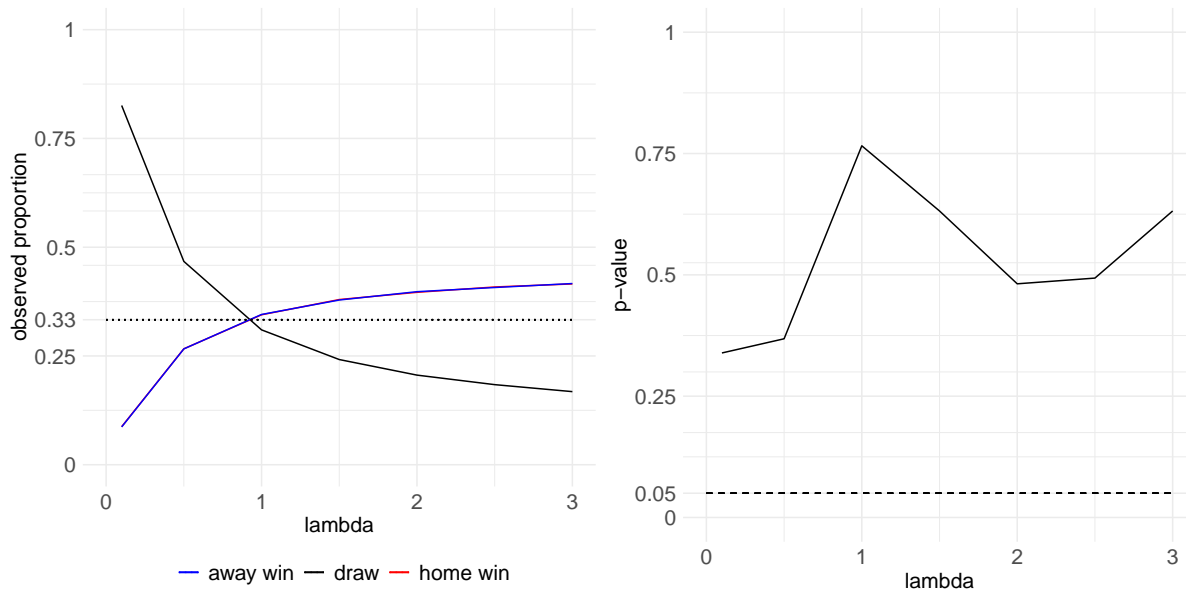}
    \caption{Results from 10000 simulated leagues under \Cref{asmp:iid} for different $\lambda$: (Left) Proportions of the three types of results; (Right) $p$-value of the multinomial goodness-of-fit tests.}
    \label{fig:simulated-samelambda}
\end{figure}

\Cref{fig:simulated-samelambda} shows that the proportion of draws decrease steadily as $\lambda$ increases. The iid assumption of the teams and the games also imply that the proportion of home wins and away wins are expected to be equal, and that is reflected in the left panel of the plot. Right panel of the figure shows the $p$-values for the multinomial goodness of fit tests based on the 10000 replications. It is evident that in all cases, the tests fail to reject the null hypothesis of every permutation being equally likely.

\begin{figure}[!hbt]
    \centering
    \includegraphics[width = \textwidth,keepaspectratio]{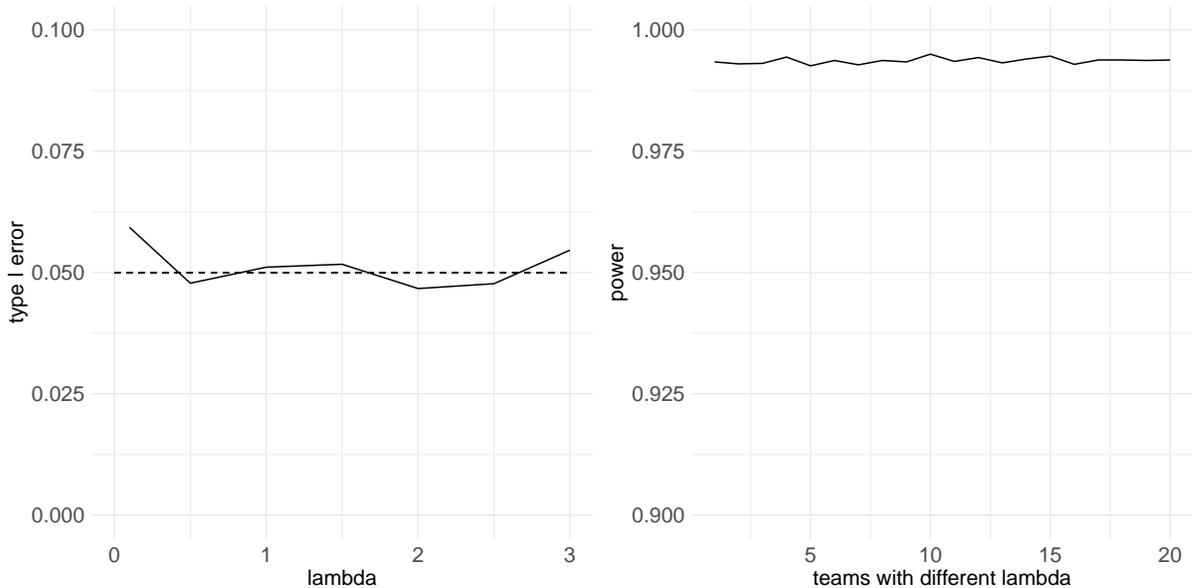}
    \caption{Results on the GBI based test: (Left) Type I error from 10000 simulated leagues under \Cref{asmp:iid} for different $\lambda$; (Right) Power from 10000 simulated leagues with some number of teams having different scoring pattern.}
    \label{fig:type1-error}
\end{figure}

Next, we wish to evaluate the effectiveness of GBI in testing whether the goal-scoring patterns of all the teams in the league are iid. We consider leagues of $n=20$ teams in this case. Once again, for different values of $\lambda$, we assume that all $X_{ijk}$ are iid $\poi(\lambda)$ and all matches are simulated accordingly. Then, the chi-squared test using GBI, as described in the previous section, is conducted. We replicate the experiment 10000 times to compute the empirical type I error of the test. These values are displayed in the left panel of \Cref{fig:type1-error}. We see that for all values of $\lambda$, the type I error remains at around 5\%. We also compute the power of the test by simulating leagues where all $X_{ijk}$ are not iid. Here, we assume that $X_{ijk}$ are independent and $X_{ijk} \sim \poi(1+\lambda_i)$, where $\lambda_i$ is a team-specific parameter. 20 different scenarios are considered in this regard. In the $k$th such scenario, $\lambda_i$ is assumed to be 0 for all but $k$ teams, randomly selected from the 20. Then, the GBI is used to detect deviation from the ideal situation. Similar to before, each experiment is repeated 10000 times and we use them to calculate the power, which is presented in the right panel of \Cref{fig:type1-error}. It is evident that the testing procedure performs very well, recording more than 99\% power even when only one team has different goal-scoring pattern. 

\section{Application}
\label{sec:results}

As mentioned above, we consider ten seasons' data from the top five leagues of Europe - Bundesliga, EPL, La Liga, Ligue 1 and Serie A. These data are collected from the public repository Datahub (link: \href{https://datahub.io/}{https://datahub.io/}). In view of \Cref{thm:lambda0-equal-prob} and the discussions following that, we begin our analysis by computing $\hat\lambda$ (see \cref{eq:lambda-mle}) for every season and every league. Corresponding theoretical probabilities of draw under the assumptions of this study and the observed proportions of drawn matches are also computed. These quantities are denoted as $d_{\hat\lambda}$ and $\hat D$, respectively. Overall, the hypothesized probabilities are close to the observed proportions of draws. There are a few cases, for example 2013-14 and 2018-19 seasons of EPL, 2010-11 season of La Liga and Ligue 1, and 2014-15 season of Serie A, when the difference between $d_{\hat\lambda}$ and $\hat D$ is more than or equal to 0.05. These results are shown in \Cref{tab:draw-summary}.

\begin{table}[!htb]
\centering
\begin{tabular}{llcccccccccc}
  \hline
League &  & \rot{2009-10} & \rot{2010-11} & \rot{2011-12} & \rot{2012-13} & \rot{2013-14} & \rot{2014-15} & \rot{2015-16} & \rot{2016-17} & \rot{2017-18} & \rot{2018-19} \\ 
  \hline
  Bundesliga & $\hat\lambda$ & 1.42 & 1.46 & 1.43 & 1.47 & 1.58 & 1.38 & 1.42 & 1.43 & 1.40 & 1.59 \\ 
   & $d_{\hat\lambda}$ & 0.25 & 0.25 & 0.25 & 0.25 & 0.24 & 0.26 & 0.25 & 0.25 & 0.25 & 0.24 \\ 
   & $\hat D$ & 0.28 & 0.21 & 0.26 & 0.26 & 0.21 & 0.27 & 0.23 & 0.24 & 0.27 & 0.24 \\ 
  \hline 
  EPL & $\hat\lambda$ & 1.39 & 1.40 & 1.40 & 1.40 & 1.38 & 1.28 & 1.35 & 1.40 & 1.34 & 1.41 \\ 
   & $d_{\hat\lambda}$ & 0.25 & 0.25 & 0.25 & 0.25 & 0.25 & 0.27 & 0.26 & 0.25 & 0.26 & 0.25 \\ 
   & $\hat D$ & 0.25 & 0.29 & 0.24 & 0.28 & 0.20 & 0.24 & 0.28 & 0.22 & 0.26 & 0.19 \\ 
  \hline 
  La Liga & $\hat\lambda$ & 1.36 & 1.37 & 1.38 & 1.44 & 1.38 & 1.33 & 1.37 & 1.47 & 1.35 & 1.29 \\ 
    & $d_{\hat\lambda}$ & 0.26 & 0.26 & 0.25 & 0.25 & 0.26 & 0.26 & 0.26 & 0.25 & 0.26 & 0.26 \\ 
   & $\hat D$ & 0.25 & 0.21 & 0.25 & 0.22 & 0.23 & 0.24 & 0.24 & 0.23 & 0.23 & 0.29 \\ 
  \hline 
  Ligue 1 & $\hat\lambda$ & 1.21 & 1.17 & 1.26 & 1.27 & 1.23 & 1.25 & 1.26 & 1.31 & 1.36 & 1.28 \\ 
   & $d_{\hat\lambda}$ & 0.28 & 0.28 & 0.27 & 0.27 & 0.27 & 0.27 & 0.27 & 0.26 & 0.26 & 0.27 \\ 
   & $\hat D$ & 0.26 & 0.34 & 0.28 & 0.28 & 0.28 & 0.23 & 0.28 & 0.25 & 0.25 & 0.29 \\ 
  \hline 
  Serie A & $\hat\lambda$ & 1.31 & 1.26 & 1.28 & 1.32 & 1.36 & 1.35 & 1.29 & 1.48 & 1.34 & 1.34 \\ 
   & $d_{\hat\lambda}$ & 0.26 & 0.27 & 0.27 & 0.26 & 0.26 & 0.26 & 0.27 & 0.25 & 0.26 & 0.26 \\ 
   & $\hat D$ & 0.27 & 0.26 & 0.29 & 0.25 & 0.24 & 0.32 & 0.25 & 0.21 & 0.22 & 0.28 \\ 
   \hline
\end{tabular}
\caption{Comparison of theoretical probabilities and the true proportions of draw for every league and every season. $\hat\lambda$ is the average number of goals, $d_{\hat\lambda}$ is the corresponding theoretical probability, and $\hat D$ is the true proportion of draws.}
\label{tab:draw-summary}
\end{table}

\begin{figure}[!htb]
    \centering
    \includegraphics[width = \textwidth,keepaspectratio]{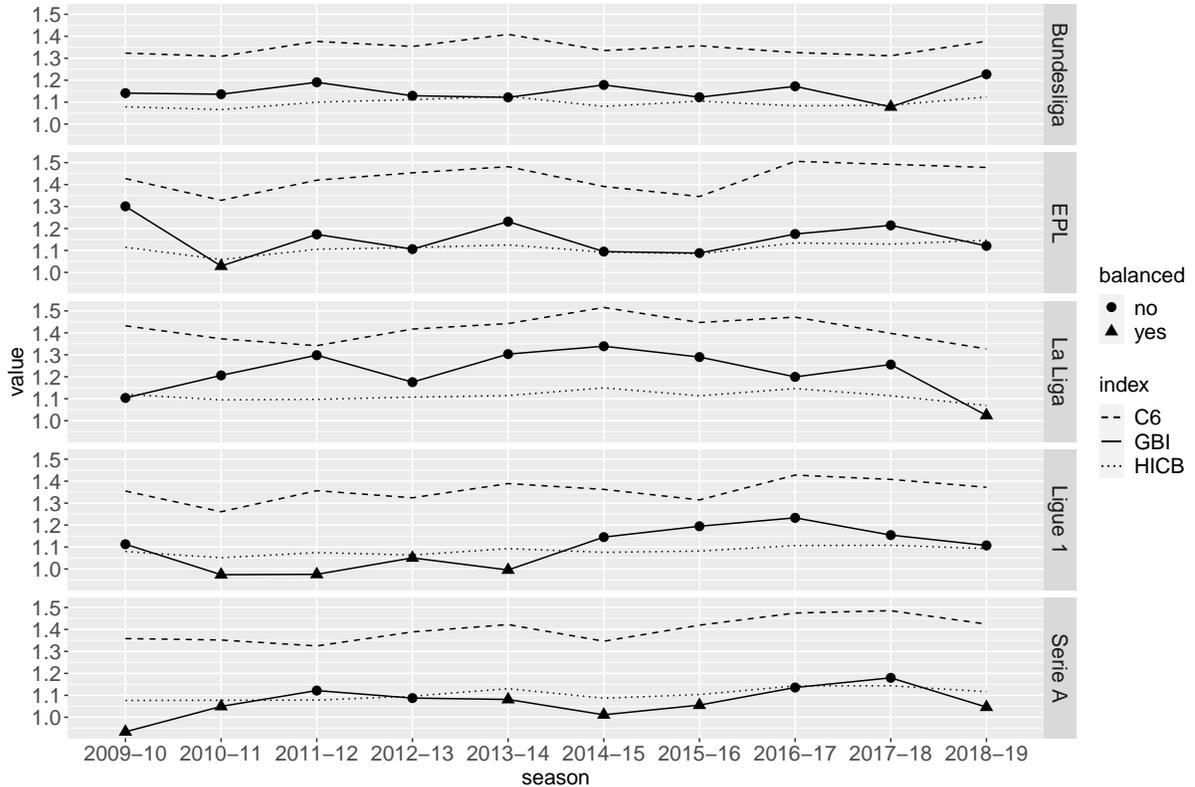}
    \caption{Three measures of competitive balance for the five leagues in ten seasons. GBI values are marked with a circle or a triangle, depending on whether the values significantly deviate from the assumption of perfect balance or not.}
    \label{fig:measures-cb}
\end{figure}

Next, the GBI values are computed for every league, across the ten seasons, and are compared against the C6 indexes and the HICB values. These quantities are displayed in \Cref{fig:measures-cb}. For the GBI values, we also test whether the league significantly deviates from the hypothetical assumption of perfect balance. In the graph, a black circle indicates that the league was not perfectly balanced in that season whereas a black triangle says the opposite. 

There are a couple of interesting aspects that arise out of this plot. Bundesliga, EPL and La Liga, barring a single season for each, have never been perfectly balanced. For Bundesliga, the only season that did not deviate significantly from the assumption of perfectly balanced was 2017/18. We find that it was a season when three teams (Schalke, Hoffenheim and Borussia Dortmund) finished on 55 points each, and the qualification for UEFA Champions League had to be determined on goal difference. Meanwhile, in the 2010-11 season of EPL, Chelsea and Manchester City finished on second and third with same number of points, and there were as many as 12 teams within a range of 10 points (39 to 49). Our approach picks that phenomena nicely and determines that this season was balanced. On the other hand, La Liga has usually observed a higher GBI value, thereby pointing to less competitiveness within the league. The only exception was the 2018-19 season, when three pairs of teams (Getafe and Sevilla, Espanyol and Athletic Bilbao, Valladolid and Celta Vigo) finished on equal points while three more teams (Real Sociedad, Real Betis, Alaves) finished on exactly same points as well.

So far as Ligue 1 is concerned, it has not observed competitive balance after the 2013-14 season. It might be interesting that this happened shortly after the huge investment on Paris Saint-Germain (PSG) by Qatar Sports Investments, the new majority shareholders of the club. Since 2013-14, PSG has topped the league in all but 2016-17 season whereas in the six Ligue 1 seasons before 2013-14, there were six different winners. Not only PSG, AS Monaco FC also spent huge amount after Russian billionaire Dmitry Rybolovlev bought two-thirds of the share and the club secured promotion back to the top division in 2013. The disparity in spending by these few clubs might have caused the earlier competitive balance go away for Ligue 1. Serie A, on the contrary, has been the most balanced league of the lot. In only four out of the ten seasons, their GBI index is significantly higher and there is no discernible pattern in the season-by-season values.

We further notice that the other two indexes also behave generally in the same directions. The overall correlation coefficient between GBI and C6 is 0.52 and the same for GBI and HICB is 0.46. C6 and HICB are more correlated with a coefficient of 0.9. The league-wise correlation coefficients for the three measures are presented in \Cref{tab:corr-measures}. 

\begin{table}[!htb]
    \centering
    \begin{tabular}{lccc}
    \hline
    League      & C6 and HICB & C6 and GBI & HICB and GBI \\
    \hline
    Bundesliga  & 0.90 & 0.31   &  0.19 \\
    EPL         & 0.94 & 0.59   &  0.59 \\
    La Liga     & 0.93 & 0.48   &  0.52 \\
    Ligue 1     & 0.91 & 0.47   &  0.61 \\
    Serie A     & 0.95 & 0.57   &  0.65 \\
    \hline
    \end{tabular}
    \caption{Correlation coefficient of the three measures for different leagues.}
    \label{tab:corr-measures}
\end{table}

\begin{figure}[!htb]
    \centering
    \includegraphics[width = \textwidth,keepaspectratio]{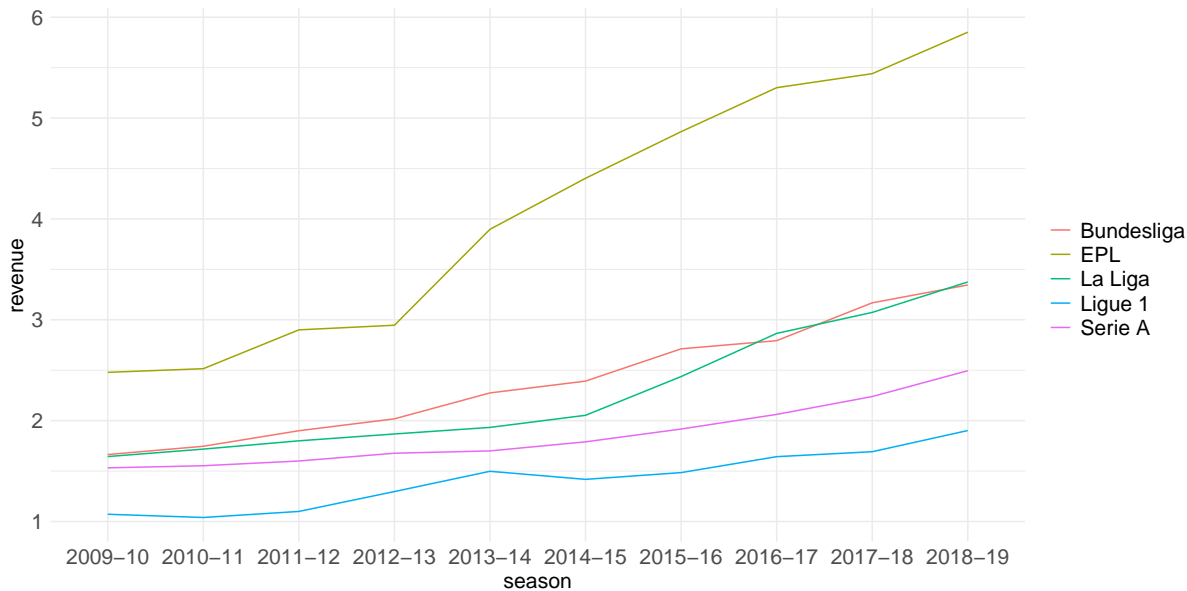}
    \caption{Season-wise revenue (in billion Euros) of the top five European leagues.}
    \label{fig:revenue}
\end{figure}

Final piece of the analysis is to analyze the relationship between revenue and the three types of measures. \Cref{fig:revenue} shows the revenues for every season for the five leagues. The growths of the overall revenues are similar for all the leagues, and that justifies the use of the linear trend term in our panel data model. It is clear that EPL has been the biggest league in terms of revenues generated over the years. Ligue 1 sits at the bottom of the list.

As mentioned before, panel data models (see \cref{eq:lm-revenue}) are used for the statistical analysis. For each of the three measures, a fixed-effect model (denoted as FE) and a random-effect model (denoted as RE) are implemented on the data. For the RE models, we use the method of \cite{swamy1972exact}. All of the results are displayed in \Cref{tab:panel}. There, $\sigma$ indicates the estimate of the idiosyncratic standard deviation of the model, that is the standard deviation of the error process for a random-effect model; and $\sigma_\alpha$ is the estimated standard deviation for the random effect.

\begin{table}[!htb]
\begin{center}
\begin{tabular}{l|cc|cc|cc}
\hline
 & FE(C6) & RE(C6) & FE(HICB) & RE(HICB) & FE(GBI) & RE(GBI) \\
\hline
(Intercept) &              & $2.24$       &              & $2.67$       &              & $2.93^{**}$  \\
            &              & $(1.72)$     &              & $(3.16)$     &              & $(0.93)$     \\
trend       & $0.21^{***}$ & $0.20^{***}$ & $0.21^{***}$ & $0.20^{***}$ & $0.21^{***}$ & $0.21^{***}$ \\
            & $(0.02)$     & $(0.02)$     & $(0.02)$     & $(0.02)$     & $(0.02)$     & $(0.02)$     \\
C6          & $-0.87$      & $-0.69$      &              &              &              &              \\
            & $(1.22)$     & $(1.23)$     &              &              &              &              \\
HICB        &              &              & $-1.57$      & $-1.27$      &              &              \\
            &              &              & $(2.88)$     & $(2.90)$     &              &              \\
GBI         &              &              &              &              & $-1.52^{*}$  & $-1.47^{*}$  \\
            &              &              &              &              & $(0.72)$     & $(0.72)$     \\
Bundesliga  & $2.45$      &                & $2.99$      &                & $3.01^{***}$ &    \\
            & $(1.61)$    &                & $(3.10)$    &                & $(0.82)$     &      \\    
EPL         & $4.18^*$      &              & $4.67$      &                & $4.67^{***}$ &    \\
            & $(1.71)$    &                & $(3.15)$    &                & $(0.83)$     &      \\    
La Liga     & $2.38$      &                & $2.89$      &                & $2.99^{**}$ &    \\
            & $(1.69)$    &                & $(3.15)$    &                & $(0.88)$     &      \\    
Ligue 1     & $1.47$      &                & $1.98$      &                & $1.94^{*}$  &    \\
            & $(1.62)$    &                & $(3.07)$    &                & $(0.79)$     &      \\    
Serie A     & $1.95$      &                & $2.46$      &                & $2.34^{***}$ &    \\
            & $(1.67)$    &                & $(3.13)$    &                & $(0.77)$     &      \\    
\hline
Adjusted $R^2$ & $0.67$      & $0.68$         & $0.67$         & $0.68$         & $0.70$         & $0.71$         \\
$\sigma$    &                & $0.39$       &                & $0.39$       &                & $0.38$       \\
$\sigma_\alpha$ &            & $0.89$       &                & $0.91$       &                & $1.05$      \\
Hausman test $p$-value &          & $0.48$         &                & $0.65$         &                & $0.73$        \\
\hline
\multicolumn{7}{l}{\scriptsize{$^{***}p<0.001$; $^{**}p<0.01$; $^{*}p<0.05$}}
\end{tabular}
\caption{Results from the panel data models. Standard errors of the coefficient estimates are given in parentheses. FE denotes a fixed effect model and RE denotes a random effect model. $\sigma$ and $\sigma_\alpha$ are the idiosyncratic standard deviation and the random effect standard deviation for the RE models.}
\label{tab:panel}
\end{center}
\end{table}

\spacingset{1} 
It is quite evident that the model using GBI performs better than the other two measures. In fact, only in the models with GBI (last two columns in the table), the coefficients corresponding to the competitive balance variable are significant. The linear trend term is always significant and those estimates are quite similar across different models. In case of the FE models, we see that the coefficients are usually not significant when we use C6 or HICB. On the contrary, they are highly significant when we use GBI, which seems to better explain the variability in the revenues.

The $p$-values from the Hausman tests between the FE and RE models are always non-significant, implying that the RE models are providing better fits for the data. Further, note that the standard deviation for the random effect is much higher than the idiosyncratic standard deviation for all RE models. All in all, RE model with GBI has proved to be the best model here.

\section{Concluding remarks}
\label{sec:discussion}

In this study, we have developed a new measure of competitive balance for a football league and have showed that it is more suitable than the existing approaches. Earlier measures of competitive balance, such as concentration ratio or Herfindahl index, were primarily developed to analyze the balance between firms in an industry; but a football league is more interesting because of some specific sets of rules to determine the ordering of the clubs. In this work, we have discussed the ideal assumptions under which a league becomes perfectly balanced when we take into account the actual rules of the leagues. To the best of our knowledge, it is the first paper to provide such a mathematical understanding. As the goal-based index we propose relies on the scorelines of the matches instead of only the outcomes, it has a natural advantage over HICB or C6 index. Moreover, solid theoretical framework behind the GBI makes it possible for us to formally test if a league is perfectly balanced or not. Simulation study confirms that the proposed test has great power.

Using the data from the top five European leagues, we also show that GBI can provide some valuable insights about the fluctuations in the competitive balance for a particular league. An interesting observation was the case of Ligue 1, where we see that the league starts to be imbalanced after PSG and Monaco got huge funding. Potentially, one can leverage the GBI to further analyze how investments or player transfers affect the competitive balance of the league. One can also hypothesize that the imbalance in a league results in significant changes in revenue sharing or broadcasting rights. A formal analysis with a bigger dataset in this regard is suitable for a great follow-up study to the current work. 

On the other hand, we have quantified the seasonal imbalance and provided relevant results. A natural extension is to develop match-level and gameweek-level measures of competitive balance; and to explore how that can be used to predict the final standing of the season. It can be possible to analyze the short-term impact of the gameweek-level competitive balance, for instance on the attendance or television viewership of every match, as well. One can also look at the competitive balance from the championship or relegation perspective and see if that can act as a significant factor in sports economics studies.

\section*{Funding source}

This research did not receive any specific grant from funding agencies in the public, commercial, or not-for-profit sectors.

\section*{Declarations of interest}

The author declares no conflict of interest.

\bibliography{references}

\section*{Appendix}

\begin{proof}[Proof of \Cref{thm:lambda0-equal-prob}]
Let us consider a game where the $i$th team is playing at home against the $j$th team. For convenience, let us use $Y_1$ and $Y_2$ to denote $X_{ij1}$ and $X_{ji0}$, respectively. First, we consider the probability of draw, i.e. $P(Y_1=Y_2|\lambda)$. Using the independence assumption and the Poisson probability mass function, 
\begin{equation}
\label{eq:prob-draw}
    P(Y_1=Y_2|\lambda) = \sum_{k=0}^\infty \frac{e^{-2\lambda}\lambda^{2k}}{(k!)^2}.
\end{equation}

Note that the modified Bessel function of the first kind of order $\alpha$ is defined by
\begin{equation}
\label{eq:modified-bessel}
    I_\alpha(x) = \sum_{k=0}^\infty \frac{1}{k!\Gamma(k+\alpha+1)}\left(\frac{x}{2}\right)^{2k+\alpha}.
\end{equation}

Clearly, \cref{eq:prob-draw} can be rewritten as $P(Y_1=Y_2|\lambda) = e^{-2\lambda}I_0(2\lambda)$. As $\lambda\to 0$, this probability goes to 1. For large $z$, $I_0(z)$ is approximately equal to $z^{-1/2}e^z/\sqrt{2\pi}$ and therefore, $P(Y_1=Y_2|\lambda) \to 0$ as $\lambda\to\infty$. Using the continuity of both terms, it is easy to argue that there exists a $\lambda_0>0$ such that the probability of draw is $1/3$. Finally, using the iid assumption, $P(Y_1>Y_2|\lambda_0) = P(Y_1<Y_2|\lambda_0)$ and that completes the proof.
\end{proof}

\begin{proof}[Proof of \Cref{thm:lambda-competitive}]
Let $Y_{ij}=X_{ij1}-X_{ji0}$. Clearly, $Y_{ij}>0$ indicates a home win for the $i$th team, $Y_{ij}<0$ indicates an away win for the $j$th team and $Y_{ij}=0$ corresponds to a draw. Since all $X_{ijk}$'s are iid $\poi(\lambda)$, $Y_{ij}$'s are iid and follow Skellam distributions (\cite{skellam1946frequency}) with parameters $(\lambda,\lambda)$. Using the modified Bessel function of the first kind (see \cref{eq:modified-bessel}), the probability mass function (pmf) of this distribution can be written as
\begin{equation}
    \label{eq:pmf-skellam}
    P(Y_{ij}=k) = e^{-2\lambda}I_{\abs{k}}(2\lambda),
\end{equation}
and the corresponding moment generating function (MGF) is
\begin{equation}
    \label{mgf:skellam}
    M(t) = \exp\left(\lambda(e^t-e^{-t})^2\right).
\end{equation}

It is evident that the distribution of $Y_{ij}$ is symmetric. Let $P(Y_{ij}=0)=e^{-2\lambda}I_0(2\lambda)$ be denoted as $d_\lambda$. Then, the probability of winning for any team in any match is $(1-d_\lambda)/2=w_\lambda$, say. Thus, the pmf of $S_{ijk}$ is 
\begin{equation}
    \label{eq:pmf-sijk}
    P(S_{ijk} = 0) = w_\lambda, \ P(S_{ijk} = 1) = d_\lambda, \ P(S_{ijk} = 3) = w_\lambda, 
\end{equation}
and $P(S_{ijk} = r) = 0$ otherwise.

Without loss of generality, consider any two teams and let $S_1$, $S_2$ be their total points in the entire league. One can write $S_1=S'_1+S_{12}$, where $S_{12}=S_{121}+S_{120}$, and $S'_1$ is the total points gathered against all but the 2nd team. Similarly, write $S_2=S'_2+S_{21}$. Under the given assumptions, it is easy to note that
\begin{equation}
    \label{eq:joint-s12}
    P(S_{12}=a,S_{21}=b) = \begin{cases} w_\lambda^2 & \text{for } a=6,b=0, \; \text{or } a=0,b=6, \\ 2w_\lambda d\lambda & \text{for } a=4,b=1, \; \text{or } a=1,b=4, \\ 2w_\lambda^2 & \text{for } a=3,b=3, \\ d_\lambda^2 & \text{for } a=2,b=2, \\ 0 & \text{otherwise}. \end{cases}
\end{equation}

Denote the support of $(S_{12},S_{21})$ by $B\subset A\times A$, with $A=\{0,1,2,3,4,6\}$. Then, 
\begin{equation}
    \label{eq:s1s2-diff-step1}
    P(S_1 > S_2) = P(S_1'-S_2' > S_{21}-S_{12}) = \sum_{(a,b)\in B} P(S_1'-S_2' > b-a)P(S_{12}=a,S_{21}=b).
\end{equation}

In view of the iid nature of every game, $S_1'$ and $S_2'$ are iid random variables. Therefore, $P(S_1'-S_2' > -r) = P(S_2'-S_1' > -r) = 1-P(S_1'-S_2' \le r)$, and thereby $P(S_1'-S_2' > -r) + P(S_1'-S_2' > r) = 1-P(S_1'-S_2' = r)$. It further implies the following.
\begin{equation}
    \label{eq:s1s2-diff-step2}
    P(S_1 > S_2) = w_\lambda^2(1 - P(S_1'-S_2' = 6)) + 2w_\lambda d\lambda(1 - P(S_1'-S_2' = 3)) + (2w_\lambda^2+d_\lambda^2)P(S_1'=S_2').
\end{equation}

A straightforward implication of the above, considering iid nature of $S_1'$ and $S_2'$, is that $P(S_1 > S_2) = P(S_1 < S_2)$. From \cref{{eq:joint-s12}}, one should also note that $P(S_{12}>S_{21})=P(S_{21}>S_{12})$.

Next, recall the definitions of $GS_1$, $GS_2$, $GD_1$, $GD_2$ from \cref{eq:overall-numbers}. Let $GS_1=GS_1'+GS_{12}$, $GS_2 = GS_2'+GS_{21}$, $GD_1=GD_1'+GD_{12}$ and $GD_2=GD_2'+GD_{21}$, defined on an identical fashion as above. It is easy to see that $GS_1'$, $GS_2'$ are iid $\poi(m\lambda)$ for $m=2n-4$, $GS_{12}$ and $GS_{21}$ are iid $\poi(2\lambda)$, $GD_{12}=GS_{12}-GS_{21}$ follows a Skellam distribution with parameters $(2\lambda,2\lambda)$, and $GD_{21}=-GD_{12}$. Thus, $P(GD_1 > GD_2) = P(GD_1'-GD_2' > GD_{21}-GD_{12}) = P(V + 2GD_{12}>0)$, where $V$ follows a Skellam distribution with parameters $(m\lambda,m\lambda)$. Clearly, both $V$ and $GD_{12}$ are symmetric around 0 and hence, simple calculations can prove that $P(GD_1 > GD_2) = P(GD_2 > GD_1)$. In a similar line, one can also show that $P(GS_1 > GS_2) = P(GS_2 > GS_1)$.

Combining the above results, we can easily argue that the final standing of the two teams is equally likely to be $(1,2)$ or $(2,1)$, if they are determined sequentially by $(S_1,S_2)$, $(GD_1,GD_2)$, $(GS_1,GS_2)$ and $(S_{12},S_{21})$. In fact, the same can be proved for any other criteria which relies on the iid assumption.

As a last piece of the proof, note that if there are two or more permutations of the teams which are not equally probable to be the final standing, then there are two teams which are not equally likely to be above each other in the league, a contradiction to what we have proved above.
\end{proof}

\begin{proof}[Proof of \Cref{thm:gbi-asymptotic}]
$X_{ijk}$, for all $i,j \in \{1,2,\hdots,n\}, k\in\{1,0\}$, are independent. Let $X_{ijk}$ be Poisson with mean $\lambda_{ijk}$. Now, consider the null hypothesis $H_0$: $\lambda_{ijk}=\lambda$ for all $i,j \in \{1,2,\hdots,n\}, k\in\{1,0\}$, against the alternate hypothesis $H_1$: $\lambda_{ijk}$'s are not all equal. The joint likelihood of $X_{ijk}$'s can be written as
\begin{equation}
    \label{eq:joint-likelihood}
    L\left(\lambda_{ijk};i,j \in \{1,2,\hdots,n\}, k\in\{1,0\}\right) = \underset{i\ne j}{\prod_{i=1}^n\prod_{j=1}^n} \prod_{k=0}^1 \frac{e^{-\lambda_{ijk}}\lambda_{ijk}^{X_{ijk}}}{X_{ijk}!}.
\end{equation}

Let $\Lambda$ be the likelihood ratio. It is easy to see that the MLE of $\lambda_{ijk}$ under $H_0$ is $\hat\lambda$ from \cref{eq:lambda-mle}, and the MLE of $\lambda_{ijk}$ under $H_0\cup H_1$ is $X_{ijk}$. Then, the standard likelihood ratio test statistic is
\begin{equation}
    \label{eq:lrt}
    -2\log\Lambda = -2 \underset{i\ne j}{\sum_{i=1}^n\sum_{j=1}^n} \sum_{k=0}^1 \left[ \left(X_{ijk} - \hat\lambda\right) + X_{ijk} \log\left(\frac{\hat\lambda}{X_{ijk}}\right) \right] = 2 \underset{i\ne j}{\sum_{i=1}^n\sum_{j=1}^n} \sum_{k=0}^1 X_{ijk} \log\left(\frac{X_{ijk}}{\hat\lambda}\right).
\end{equation}

Take $f(x)=x\log(x/x')$ for a fixed $x'$. The Taylor series expansion of $f(x)$ about $x'$ is
\begin{equation*}
    f(x) = (x-x') + \frac{1}{2x'}(x-x')^2 + \delta,
\end{equation*}
where $\delta$ is negligible under $H_0$ and for a large sample, which is ensured considering $N=O(n^2)$. Thus, under that scenario, the likelihood ratio test statistic can be approximated as
\begin{equation}
    \label{eq:lrt-taylor}
    -2\log\Lambda = 2 \underset{i\ne j}{\sum_{i=1}^n\sum_{j=1}^n} \sum_{k=0}^1 \left[ \left(X_{ijk}-\hat\lambda\right) + \frac{\left(X_{ijk}-\hat\lambda\right)^2}{2\hat\lambda}\right].
\end{equation}

Since the first term above adds up to 0, it is easy to see that $-2\log\Lambda$ is asymptotically equivalent with $(2N-1)\mathrm{GBI}$. Finally, the required result follows from the well known theory about likelihood ratio tests that $-2\log\Lambda \sim \chi^2$ with degrees of freedom $2N-1$. 
\end{proof}

\end{document}